\pdfoutput=1
\documentclass[prl,twocolumn,showpacs,amsmath,amssymb,floatfix]{revtex4-1}

\usepackage{graphicx}  
\usepackage{dcolumn}   
\usepackage{bm}        
\usepackage{amssymb}   
\usepackage{amsmath}
\usepackage{mathrsfs}
\usepackage{epigraph}
\usepackage{braket}
\usepackage{gensymb}
\usepackage{lmodern}
\usepackage{lipsum}
\usepackage{marvosym}
\usepackage{mathdots}
\usepackage{ tipa }
\usepackage{bbold}
\usepackage{dsfont}
\usepackage{esint}
\usepackage{xcolor}
\usepackage{appendix}
\usepackage{natbib}
\usepackage[colorlinks=true, linkcolor=blue, urlcolor=blue,citecolor=blue]{hyperref}

\DeclareMathOperator{\sech}{sech}

\begin{document}

\title{Spintronics meets nonadiabatic molecular dynamics: Geometric spin torque and damping on noncollinear classical magnetism due to electronic open quantum system} 

\author{Utkarsh Bajpai}
\affiliation{Department of Physics and Astronomy, University of Delaware, Newark, DE 19716, USA}

\author{Branislav K. Nikoli\'{c}}
\affiliation{Department of Physics and Astronomy, University of Delaware, Newark, DE 19716, USA}

\begin{abstract}
We analyze a quantum-classical hybrid system of  steadily precessing slow classical localized magnetic moments, forming a head-to-head domain wall,  embedded into an open quantum system of fast nonequilibrium electrons. The electrons reside within a metallic wire connected to macroscopic reservoirs. The model  captures the essence of dynamical noncollinear and noncoplanar magnetic textures in spintronics, while making it  possible to obtain the {\em exact} time-dependent nonequilibrium density matrix of electronic system and split it into four contributions. The Fermi surface contribution generates dissipative (or damping-like in spintronics terminology) spin torque on the moments, and one of the two Fermi sea contributions generates {\em geometric} torque dominating in the adiabatic regime. When the coupling to the reservoirs is reduced, the geometric torque is the only nonzero contribution. Locally it has both nondissipative (or field-like in spintronics terminology) and damping-like components, but with the sum of latter being zero, which act as the counterparts of geometric magnetism force and  electronic friction in nonadiabatic molecular dynamics. Such {\em current-independent} geometric torque is {\em absent} from widely used micromagnetics or atomistic spin dynamics modeling of magnetization dynamics based on the Landau-Lifshitz-Gilbert  equation, where previous analysis of Fermi surface-type torque has severely underestimated its magnitude. 
\end{abstract}

\maketitle

One of the most fruitful applications of geometric (or Berry) phase~\cite{Berry1984} concepts is encountered in quantum-classical hybrid systems  where separation of time scales makes it possible to consider fast quantum degrees of freedom interacting with the slow classical ones~\cite{Berry1993,Zhang2006}. The  amply studied example of this kind are fast  electrons interacting~\cite{Min2014,Requist2016} with slow nuclei in molecular dynamics (MD)~\cite{Ryabinkin2017,Dou2020,Dou2018,Lu2019} problems of physics, chemistry and biology. The parameters driving adiabatic evolution of quantum subsystem, with characteristic frequency smaller that its level spacing, are nuclear coordinates elevated to the status of dynamical variables. The electronic system then develops geometric phase in states evolving out of an instantaneous energy eigenstate, while also acquiring shifts in the  energy levels. Conversely, nuclei experience  forces due to {\em back-action} from electrons. The simplest force is the {\em adiabatic} Born-Oppenheimer (BO) force~\cite{Min2014,Requist2016} which depends only on the coordinates of the nuclei, and it is associated with electronic adiabatic potential surfaces~\cite{Ryabinkin2017,Dou2020}. Even small violation of BO approximation leads to additional forces---the {\em first nonadiabatic} correction generates forces linear in the velocity of the nuclei, and being Lorentz-like they are dubbed~\cite{Berry1993,Campisi2012} ``geometric magnetism.'' The ``magnetism'' is not a not a real magnetic field, but an emergent geometrical property of the Hilbert space~\cite{Kolodrubetz2017}, and akin to the true Lorentz force, the emergent geometric force is {\em nondissipative}. 

\begin{figure}
	\includegraphics[width=8.5cm]{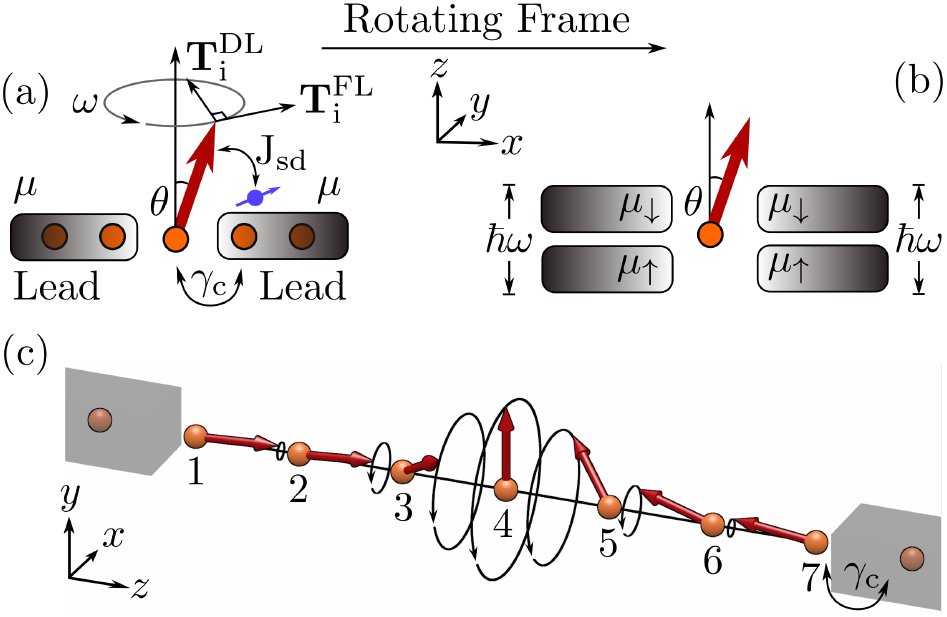}
	\caption{(a) Schematic view of a two-terminal system where a single classical 
		LMM, precessing steadily with frequency $\omega$ and cone angle $\theta$, interacts with an open quantum system of conduction electron spins. The electrons hop along 1D infinite tight-binding chain which terminates into the left and right macroscopic reservoirs kept at the same chemical potential $\mu$. Panel (c) depicts 7 LMMs, $\bold{M}_1$--$\bold{M}_7$ forming a head-to-head Bloch domain wall, which precess with the same frequency but are {\em noncollinear} and {\em noncoplanar}. Both (a) and (c) can be mapped in the rotating frame to a time-independent four-terminal system in (b) with an effective bias voltage $\hbar \omega/e$ between the left or right pair of leads.}
	\label{fig:fig1}
\end{figure}

Additional forces appear upon making the quantum system open by coupling it to a thermal bath~\cite{Campisi2012,Gaitan1998} (usually modeled as an infinite set of harmonic oscillators~\cite{Whitney2003}) or to macroscopic reservoirs of particles~\cite{Thomas2012}. In the latter case, one can also introduce chemical potential difference between the reservoirs to drive particle flux (i.e., current) through the  quantum system which is, thereby, pushed out of equilibrium~\cite{Bode2011,Thomas2012,Lu2012,Hopjan2018,Chen2019}. In both equilibrium and nonequilibrium cases, the energy spectrum of the quantum system is transformed into a continuous one, and frictional forces~\cite{Campisi2012,Bode2011,Thomas2012,Lu2012,Dou2018,Lu2019,Dou2017,Hopjan2018,Chen2019} linear in the velocity of the nuclei become possible.  Also, due to continuous spectrum, adiabaticity criterion has to be replaced by a different one~\cite{Thomas2012}. Stochastic forces also appear, both in equilibrium and in nonequilibrium, where in the former case~\cite{Campisi2012,Gaitan1998} they are due to fluctuations at finite temperature while in the latter case they include additional contribution from nonequilibrium noise~\cite{Bode2011,Thomas2012,Lu2012}. Finally, specific to nonequilibrium is the emergence of nonconservative forces~\cite{Bode2011,Thomas2012,Lu2012,Hopjan2018,Chen2019}. The derivation of all of these forces is achieved by computing nonadiabatic corrections to the density matrix (DM)~\cite{Campisi2012,Gaitan1998,Bode2011,Thomas2012,Lu2012,Hopjan2018,Chen2019}. This yields a non-Markovian stochastic Langevin equation, with nonlocal-in-time kernel describing memory effects~\cite{Farias2009}, as the most general~\cite{Lu2012,Chen2019} equation for nuclei in nonadiabatic MD. 

The analogous problem exists in spintronics, where the fast quantum system is comprised of conduction electron spins and slow classical system is comprised of localized-on-atoms spins and associated localized magnetic moments (LMMs) described by unit vectors $\mathbf{M}_i(t)$. The dynamics of LMMs is  accounted by the Landau-Lifshitz-Gilbert (LLG) type of equation~\cite{Evans2014} 
\begin{eqnarray}\label{eq:llg}
\frac{\partial \mathbf{M}_i}{dt} & = & -g \mathbf{M} \times \mathbf{B}^\mathrm{eff}_i + \lambda \mathbf{M}_i \times \frac{\partial \mathbf{M}_i}{\partial t} \nonumber \\
&& + \frac{g}{\mu_M}\left( \mathbf{T}_i \left[I^{S_\alpha}_\mathrm{ext}\right] + \mathbf{T}_i\left[\partial \mathbf{M}_i/\partial t \right] \right).  
\end{eqnarray}
This includes phenomenological Gilbert damping, whose parameter $\lambda$ can be measured or independently calculated~\cite{Starikov2010} by using electronic Hamiltonian with spin-orbit coupling and impurities. It can also include Slonczewski spin-transfer torque (STT) term $\mathbf{T}_i \left[ I^{S_\alpha}_\mathrm{ext} \right]$ due to externally supplied spin current $I^{S_\alpha}_\mathrm{ext}$. The STT is a phenomenon~\cite{Ralph2008} in which spin angular momentum of conduction electrons is transferred to local magnetization not aligned with electronic spin-polarization. Finally, some analyses ~\cite{Zhang2004,Zhang2009,Tatara2019}  also consider  {\em current-independent}  torque $\mathbf{T}_i[\partial \mathbf{M}_i/\partial t]$ as a back-action of electrons pushed out of equilibrium by time-dependent $\mathbf{M}_i(t)$. Nevertheless, 
such effects have been deemed negligible~\cite{Zhang2004,Kim2012} or easily absorbed into Eq.~\eqref{eq:llg} by renormalizing $g$ and $\lambda$~\cite{Zhang2004}. Here $g$ is the gyromagnetic ratio; $\mathbf{B}^{\rm eff}_i = - \frac{1}{\mu_M} \partial \mathcal{H} /\partial \mathbf{M}_i$ is the effective magnetic field as the sum of external field, field due to interaction with other LMMs and magnetic anisotropy field in the classical Hamiltonian $\mathcal{H}$ of LMMs; and $\mu_M$ is the magnitude of LMM~\cite{Evans2014}.  

\begin{figure}
	\includegraphics[width=8.5cm]{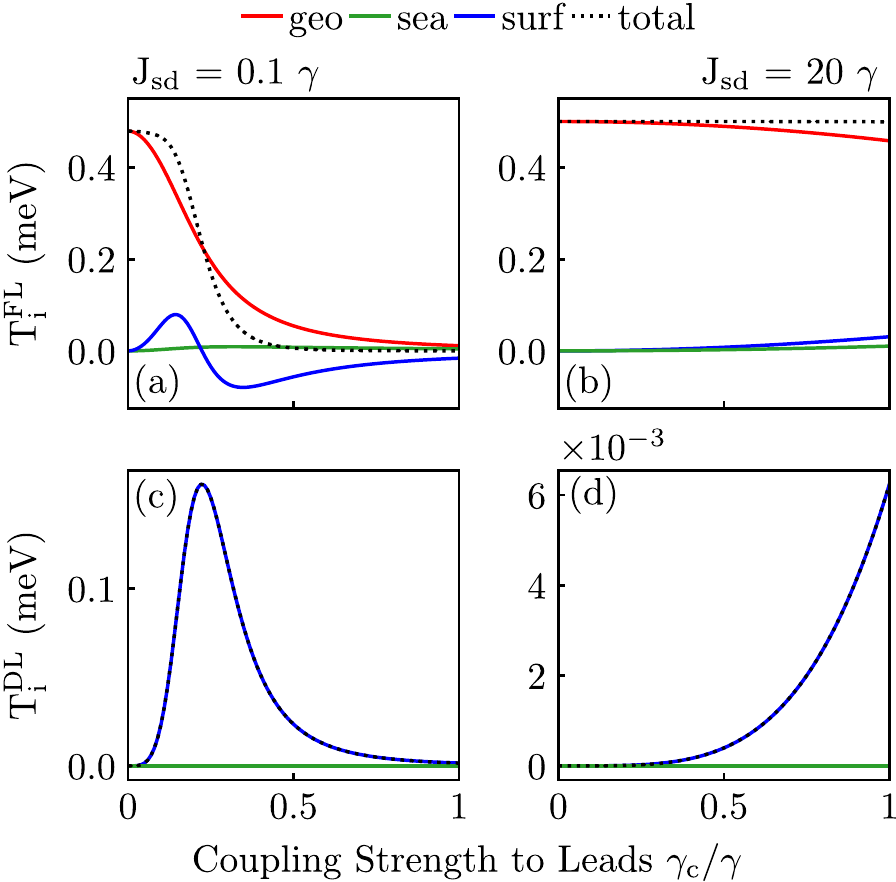}
	\caption{The FL and DL components [Fig.~\eqref{eq:llg}] of three spin torques contributions in Eq.~\eqref{eq:stt} exerted by nonequilibrium spin density of electrons onto a single localized precessing magnetic moment in the setup of Fig.~\ref{fig:fig1}(a) as a function of coupling to the leads. Black dotted line is the sum of the three torques. In panels (a) and (c) \mbox{$J_{sd} = 0.1 \gamma$}, while in panels (b) and (d) \mbox{$J_{sd} = 20 \gamma$} ensures perfectly adiabatic regime~\cite{Stahl2017}, $J_{sd}/\hbar\omega \gg 1$, for the chosen precession frequency $\hbar \omega=0.001 \gamma$.}
	\label{fig:fig2}
\end{figure}

\begin{figure*}
	\includegraphics[width=18cm]{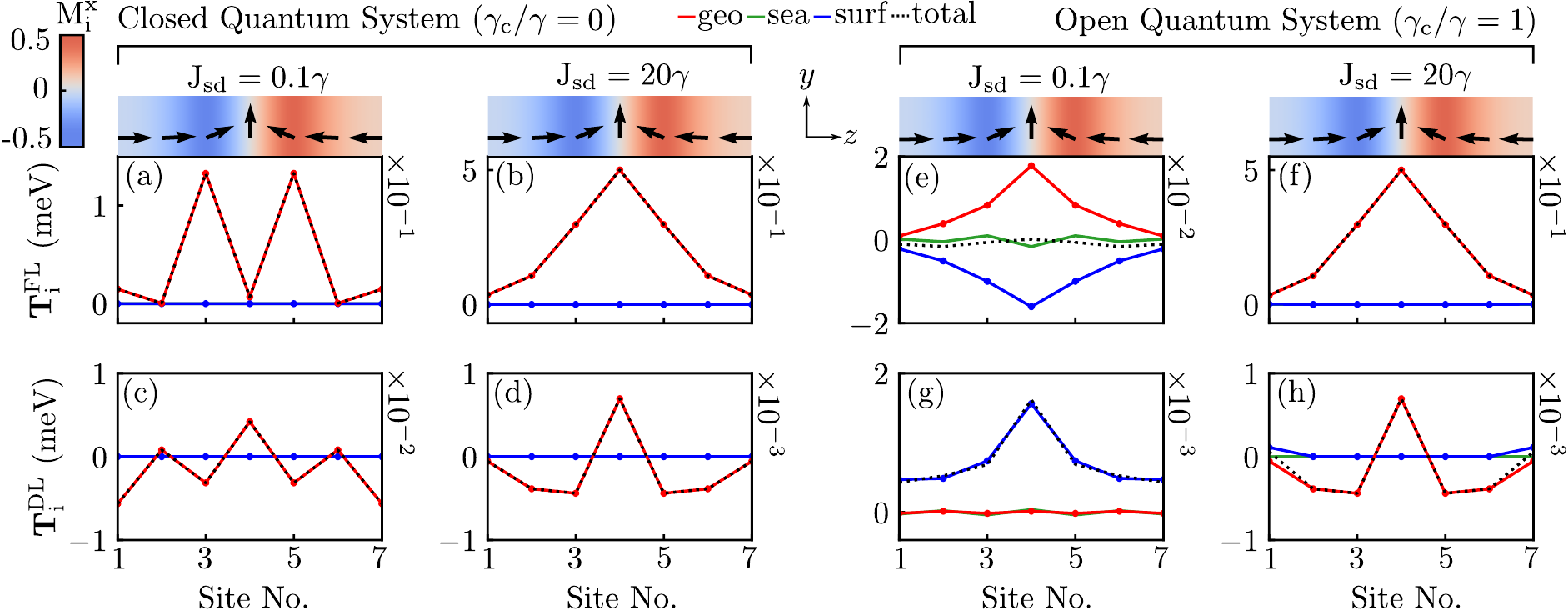}
	\caption{Spatial profile of FL and DL components of $\mathbf{T}_i^\mathrm{geo}$, $\mathbf{T}_i^\mathrm{sea}$ and $\mathbf{T}_i^\mathrm{surf}$ spin torques on precessing LMMs depicted in  Fig.~\ref{fig:fig1}(c) for closed or open electronic quantum system and for two different values of  $J_{sd}$. Insets on the top of each row mark positions and static configuration of LMMs within the Bloch DW, with their $x$-component depicted by the colorbar next to panel (a).}
	\label{fig:fig3}
\end{figure*}

The STT vector, $\mathbf{T}=\mathbf{T}^\mathrm{FL} + \mathbf{T}^\mathrm{DL}$, can be decomposed [Fig.~\ref{fig:fig1}(a)] into: ({\em i}) even under  time-reversal or field-like (FL) torque,  which affects precession of LMM around $\mathbf{B}^{\rm eff}_i$; and  ({\em ii}) odd under time-reversal or damping-like (DL) torque, which either enhances the Gilbert damping by pushing LMM toward $\mathbf{B}^{\rm eff}_i$ or competes with Gilbert term as ``antidamping.'' For example, negative values of $T^\mathrm{DL} = \mathbf{T}^\mathrm{DL} \cdot \mathbf{e}_\mathrm{DL}$ in Figs.~\ref{fig:fig2} and ~\ref{fig:fig3},  where $\mathbf{e}_\mathrm{DL}=(\bold{M}_i \times \partial \bold{M}_i/\partial t)|\bold{M}_i \times \partial \bold{M}_i/\partial t|^{-1}$, means that $\mathbf{T}^\mathrm{DL}$ vector points away from the axis of precession which is antidamping action. Similarly, $T^\mathrm{FL} = \mathbf{T}^\mathrm{FL} \cdot \mathbf{e}_\mathrm{FL}$, where $\mathbf{e}_\mathrm{FL}=(\partial \bold{M}_i/\partial t)|\partial \bold{M}_i/\partial t|^{-1}$, is plotted in Figs.~\ref{fig:fig2} and ~\ref{fig:fig3}.

%

%

The current-driven STT $\mathbf{T}_i \left[ I^{S_\alpha}_\mathrm{ext} \right]$ acts as the counterpart of nonconservative force in nonadiabatic MD. The Gilbert damping plus  current-independent torque $\mathbf{T}_i[\partial \mathbf{M}_i/\partial t]$ appear  as the counterpart of electronic friction~\cite{Bode2011,Thomas2012,Lu2012,Dou2018,Lu2019,Dou2017,Hopjan2018,Chen2019},  but Gilbert damping requires agents~\cite{Starikov2010} other than electrons alone considered in nonadiabatic MD. Thus, the {\em geometric} torque and damping, as counterparts of geometric magnetism force~\cite{Berry1993} and friction~\cite{Campisi2012}, are {\em absent} from standard modeling of classical magnetization dynamics. Geometric torque has been added {\em ad hoc} into the LLG equation applied to specific problems, such as spin waves  within bulk magnetic materials~\cite{Wen1988,Niu1998,Niu1999}. A recent study~\cite{Stahl2017} of a single classical LMM embedded into a closed (i.e., finite length one-dimensional wire) electronic quantum system finds that nonequilibrium electronic spin density always generates geometric torque, even in perfectly adiabatic regime where electron-spin/LMM interaction is orders of magnitude larger than the characteristic frequency of LMM dynamics. It acts as a purely FL torque causing  anomalous frequency of precession that is higher than the Larmor frequency. By retracing the same steps~\cite{Thomas2012,Bode2011} in the derivation of the stochastic  Langevin equation for electron-nuclei system connected to macroscopic reservoirs, Ref.~\cite{Bode2012} derived the stochastic LLG equation~\cite{Onoda2006,Nunez2008,Fransson2008,Hurst2020}  for a single LMM embedded into an open electronic system out of equilibrium. The novelty in this derivation is  damping, present even in the absence of traditional spin-flip relaxation mechanisms~\cite{Zhang2004,Tatara2019}, while the same conclusion about geometric torque changing only the precession frequency of LMM has been reached (in some regimes, geometric phase can also affect the stochastic torque~\cite{Shnirman2015}).  However, single LMM is a rather special case, which is illustrated in Fig.~\ref{fig:fig1}(a) and revisited in Fig.~\ref{fig:fig2}, and the most intriguing situations in spintronics involve dynamics of noncollinear textures of LMMs. This is exemplified by  current- or magnetic-field driven dynamics of domain walls (DWs) and skyrmions~\cite{Tatara2019,Hurst2020,Petrovic2018,Bajpai2019a,Petrovic2019,Bostrom2019,Woo2017} where a much richer panoply of back-action effects from fast  electronic system can  be expected. 

\begin{figure}
	\includegraphics[scale=0.9]{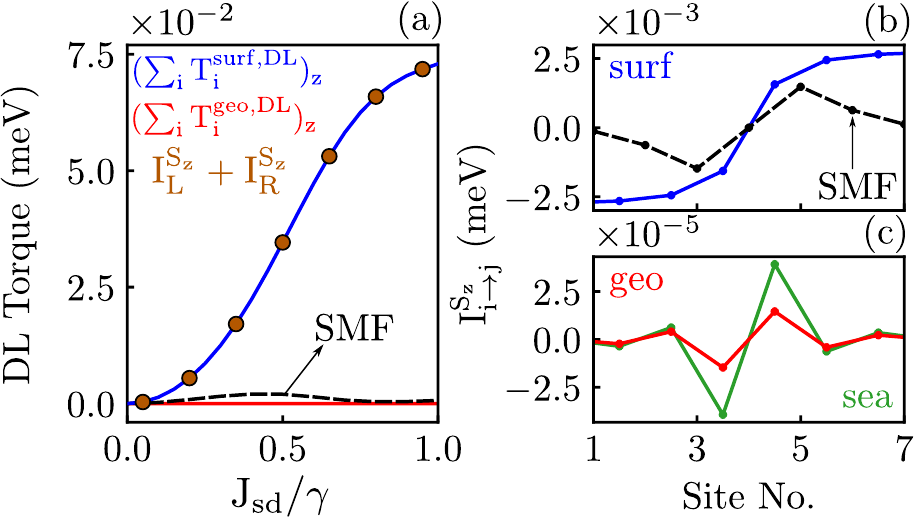}
	\caption{(a) The $z$-component of total DL torques which act on DW in Fig.~\ref{fig:fig1}(c) as a function of $J_{sd}$ for $\gamma_c=\gamma$. Circles show that sum of spin currents pumped into the leads matches $\left(\sum_i \mathbf{T}_i^\mathrm{surf,DL} \right)_z \equiv I_\mathrm{L}^{S_z} + I_\mathrm{R}^{S_z}$.  Panel (b) and (c), which correspond to Fig.~\ref{fig:fig3}(g), show spatial profile of local spin currents $I_{i \rightarrow j}^{S_z}$ pumped between sites $i$ and $j$ for $J_{sd}=0.1 \gamma$, with their sum being identically {\em zero} in panel (c). Dashed black line in panels (a) and (b) is pumped local spin current  by SMF~\cite{Zhang2009,Kim2012}, $I_\mathrm{SMF}^{S_z}(x)=\frac{g\mu_B \hbar G_0}{4e^2} [ \partial {\bf M}(x, t)/\partial t \times \partial {\bf M}(x, t)/\partial x ]_z$, where $G_0 = G^\uparrow + G^\downarrow$ is the total conductivity.}
	\label{fig:fig4}
\end{figure}

In this Letter, we analyze an exactly solvable model of seven steadily precessing LMMs, $\mathbf{M}_1(t)$--$\mathbf{M}_7(t)$ [Fig.~\ref{fig:fig1}(c)], which are noncollinear and noncoplanar and embedded into a one-dimensional (1D) infinite wire hosting conduction electrons.  The model can 
be viewed as a segment of dynamical noncollinear magnetic texture, and it directly describes magnetic field-driven~\cite{Woo2017} head-to-head Bloch DW~\cite{McMichael1997} but without allowing it to propagate~\cite{Woo2017,Petrovic2019}. Its simplicity  makes it exactly solvable---we fins the {\em exact time-dependent DM} via the nonequilibrium Green function (NEGF) formalism~\cite{Stefanucci2013}  and analyze its contributions in different regimes of the ratio $J_{sd}/\hbar \omega$ of  $sd$ exchange interaction $J_{sd}$~\cite{Zhang2004} between electron spin and LMM and frequency of precession $\omega$. In both Figs.~\ref{fig:fig1}(a) and \ref{fig:fig1}(c), the electronic subsystem is an {\em open quantum system} and, although no bias voltage is applied between the macroscopic reservoirs, it is pushed into  the {\em nonequilibrium} state by the dynamics of LMMs. For example, electronic quantum Hamiltonian becomes time-dependent due to $\mathbf{M}_1(t)$ [Fig.~\ref{fig:fig1}(a)] or $\mathbf{M}_1(t)$--$\mathbf{M}_7(t)$ [Fig.~\ref{fig:fig1}(c)], which leads to pumping~\cite{Tatara2019,Chen2009,Tserkovnyak2005} [Fig.~\ref{fig:fig4}(b),(c)] of spin current locally within the DW region, as well as into  the leads [Fig.~\ref{fig:fig4}(a)]. Pumping of charge current will also occur if the left-right symmetry of the device is broken statically~\cite{Chen2009} or dynamically~\cite{Bajpai2019}.

The electrons are modeled on an infinite tight-binding (TB) clean chain with Hamiltonian in the lab frame 
\begin{equation}\label{eq:hamiltonian}
\hat{H}_\mathrm{lab}(t) = -\gamma \sum_{\langle ij\rangle }  \hat{c}_i^\dagger\hat{c}_j
 -J_{sd}\sum_{i}\hat{c}_i^\dagger \hat{\bm \sigma} \hat{c}_i \cdot \bold{M}_i(t).
\end{equation}
Here $\hat{c}_i^\dagger = (\hat{c}_{i\uparrow}^\dagger, \hat{c}_{i\downarrow}^\dagger)$ and $\hat{c}_{i\sigma}^\dagger$ ($\hat{c}_{i\sigma}$) creates (annihilates) an electron of spin $\sigma=\uparrow, \downarrow$ at site $i$. The nearest-neighbor hopping \mbox{$\gamma = 1$ eV} sets the unit of energy. The active region  in Figs.~\ref{fig:fig1}(a) or ~\ref{fig:fig1}(c) consists of one or seven sites, respectively, while the rest of  infinite TB chain is taken into account as the left (L) and the right (R) semi-infinite leads described by the same Hamiltonian in Eq.~\eqref{eq:hamiltonian}, but with $J_{sd}=0$. The hopping between the leads and the active region is denoted  by $\gamma_c$. The leads terminate at infinity into the macroscopic particle reservoirs with identical chemical potentials $\mu_\mathrm{L}=\mu_\mathrm{R}=E_F$ due to assumed absence of bias voltage, and $E_F=0$ is chosen as the Fermi energy. In contrast to traditional analysis in spintronics~\cite{Zhang2004,Tatara2019}, but akin to Refs.~\cite{Stahl2017,Bode2012}, Hamiltonian in Eq.~\eqref{eq:hamiltonian} {\em does not} contain  any spin-orbit or impurity terms as generators of spin-flip relaxation.

The spatial profile of Bloch DW is given by $M_i^x=-\sech[(h_\mathrm{DW}-z_i)/W]\tanh[(Z_\mathrm{DW}-z_i)]$, $M_i^y=\sech^2[(Z_\mathrm{DW}-z_i)/W]$ and $M_i^z=\tanh[(Z_\mathrm{DW}-z_i)/W]$, where $Z_\mathrm{DW}=4$ and $W=0.9$. Instead of solving LLG equations [Eq.~\eqref{eq:llg}] for $\bold{M}_1(t)$--$\bold{M}_7(t)$, we impose a solution  where LMMs  precess steadily around the $z$-axis: $M_i^x(t)  = \sin \theta_i \cos(\omega t + \phi_i)$;  $M_i^y(t) = \sin \theta_i \sin(\omega t + \phi_i)$; and $M_i^z(t)  = \cos \theta_i$. Using a unitary transformation into the rotating frame (RF), the Hamiltonian in Eq.~\eqref{eq:hamiltonian} becomes time-independent~\cite{Chen2009,Tatara2019}, $\hat{H}_\mathrm{RF} = \hat{U}^\dagger(t) \hat{H}_\mathrm{lab}(t) \hat{U}(t) -i \hbar \hat{U}^\dagger \partial \hat{U}/\partial t = \hat{H}_\mathrm{lab}(t=0) - \hbar\omega\hat{\sigma}_\alpha/2$, with LMMs frozen at $t=0$ configuration from the lab. The unitary operator is \mbox{$\hat{U}(t) = \exp(-i\omega t \hat{\sigma}_\alpha/2)$} for $\alpha$-axis of rotation. In the RF, the original two-terminal Landauer setup for quantum transport in Figs.~\ref{fig:fig1}(a) and ~\ref{fig:fig1}(c) is mapped, due to  $\hbar\omega \hat{\sigma}_\alpha/2$ term, onto an effective four-terminal setup~\cite{Chen2009}  [illustrated for single LMM in Fig.~\ref{fig:fig1}(b)]. Each of its four leads is an effective half-metal ferromagnet which accepts only one spin species, $\uparrow$ or $\downarrow$ along the $\alpha$-axis, and effective dc bias voltage $\hbar\omega/e$ acts between L or R pair of leads.

In the RF, the presence of the leads and macroscopic reservoirs can be taken into account exactly using steady-state NEGFs~\cite{Stefanucci2013} which depend on time difference $t-t'$ and energy $E$ upon Fourier transform. Using the retarded, $\hat{G}(E)$, and the lesser, $\hat{G}^<(E)$, Green functions (GFs), we  find the exact nonequilibrium DM of electrons in the RF, \mbox{$\hat{\rho}_\mathrm{RF} = \frac{1}{2\pi i}\int dE \, \hat{G}^<(E)$}. Here the two GFs are related by the Keldysh equation, $\hat{G}^<(E)=\hat{G}(E)\hat{\Sigma}^<(E)\hat{G}^\dagger(E)$, where $\hat{\Sigma}^<(E)$ is the lesser self-energy~\cite{Stefanucci2013} due to semi-infinite leads and $\hat{G}(E)=[E - \hat{H}_\mathrm{RF} -\hat{\Sigma}(E, \hbar\omega) ]^{-1}$ with  $\hat{\Sigma} (E, \hbar\omega) = \sum_{p=\mathrm{L,R}, \sigma=\uparrow, \downarrow} \hat{\Sigma}_p^\sigma(E-Q_\alpha^\sigma \hbar\omega)$ being the sum of retarded self-energies for each of the four leads $p$, $\sigma$ in RF. We use shorthand notation $Q_p^\uparrow = -1/2$ and  $Q_p^\downarrow = +1/2$. Since typical frequency of magnetization dynamics is \mbox{$\hbar\omega \ll E_F$}, we can expand~\cite{Mahfouzi2013} $\hat{\rho}_\mathrm{RF}$ in small $\hbar\omega/E_F$ and then transform it back to the lab frame, $\hat{\rho}_\mathrm{lab}(t) = \hat{U}(t)\hat{\rho}_\mathrm{RF} \hat{U}^\dagger(t)$ to obtain $\hat{\rho}_\mathrm{lab}(t) = \hat{\rho}^\mathrm{ad}_t + \hat{\rho}_\mathrm{geo}(t) + \hat{\rho}_\mathrm{sea}(t) +  \hat{\rho}_\mathrm{surf}(t)$ where:
\begin{subequations}\label{eq:dm_exp}
\begin{eqnarray}
\hat{\rho}^\mathrm{ad}_t & = & -\frac{1}{\pi} \hat{U} \int\limits_{-\infty}^{+\infty} \! dE  \mathrm{Im}\hat{G}_0 f(E)\hat{U}^\dagger, \label{eq:dm_ad} \\
\hat{\rho}_\mathrm{geo}(t) &  = & \frac{1}{\pi} \hat{U}  \int\limits_{-\infty}^{+\infty} \! dE \mathrm{Im}\bigg[\hat{G}_0
\bigg( i\hbar \hat{U}^\dagger\frac{\partial \hat{U}}{\partial t} \bigg) \hat{G}_0 \bigg] f(E) \hat{U}^\dagger, \label{eq:dm_geo} \\
\hat{\rho}_\mathrm{sea}(t) &  = & -\frac{\hbar\omega}{2\pi} \hat{U} \sum_p \int\limits_{-\infty}^{+\infty} \! dE \mathrm{Im}\bigg[\hat{G}_0
\bigg( \frac{\partial \hat{\Sigma}^\uparrow_p}{\partial E}
- \frac{\partial \hat{\Sigma}^\downarrow_p}{\partial E} \bigg) \hat{G}_0 \bigg] \nonumber \\
&& \times f(E)\hat{U}^\dagger, 
\label{eq:dm_sea} \\
\hat{\rho}_\mathrm{surf}(t) & = & \frac{\hbar\omega}{4\pi}  \hat{U}  \sum_{p} \int\limits_{-\infty}^{+\infty} \! dE \hat{G}_0  (\hat{\Gamma}^\uparrow_p - \hat{\Gamma}_p^\downarrow) \hat{G}_0^\dagger \frac{\partial f}{\partial E} \hat{U}^\dagger. \label{eq:dm_surf}
\end{eqnarray}
\end{subequations}
We confirm by numerically exact calculations~\cite{Petrovic2018} that thus obtained $\hat{\rho}_\mathrm{lab}(t)$ is identical to $\hbar G^<(t,t)/i$ computed in the lab frame. Here \mbox{$\hat{G}_0(E) = [E - \hat{H}_\mathrm{RF} - \hat{\Sigma}(E, 0)]^{-1}$} is $\hat{G}(E)$ with $\hbar\omega = 0$;  $\hat{\Gamma}_p^{\sigma}(E) = i[\hat{\Sigma}_p^{\sigma}(E) - \hat{\Sigma}^{\sigma}_p(E)^\dagger]$ is the level broadening matrix due the leads; and  $f_p^\sigma(E) = f(E - [E_F + Q_\alpha^\sigma\hbar\omega])$ is the the Fermi function of  macroscopic reservoir $p$, $\sigma$ in the RF.

The total nonequilibrium spin density, $\langle \hat{\mathbf{s}}_i \rangle(t)=\mathrm{Tr}[\hat{\rho}_\mathrm{lab}(t) |i\rangle \langle i| \otimes \hat{\bm \sigma}]=\langle \hat{\mathbf{s}}_i \rangle^\mathrm{ad}_t  + \langle \hat{\mathbf{s}}_i \rangle_\mathrm{geo}(t) + \langle \hat{\mathbf{s}}_i \rangle_\mathrm{sea}(t) + \langle \hat{\mathbf{s}}_i \rangle_\mathrm{surf}(t)$, has the corresponding four contributions from DM contributions in Eq.~\eqref{eq:dm_exp}. Here  $\langle \hat{\mathbf{s}}_i \rangle^\mathrm{ad}_t$ is the equilibrium expectation value at an instantaneous time $t$ which defines 
`adiabatic spin density'~\cite{Zhang2004,Tatara2019,Niu1998,Niu1999,Stahl2017}. It is computed using  $\hat{\rho}^\mathrm{ad}_t$ as the grand canonical equilibrium DM  expressed via the frozen (adiabatic) retarded GF~\cite{Thomas2012,Bode2011,Bode2012},  \mbox{$\hat{G}_t(E)=[E-\hat{H}_t-\hat{\Sigma}]^{-1}$}, for instantaneous configuration of $\mathbf{M}_i(t)$ while assuming $\partial \mathbf{M}_i/\partial t = 0$ [subscript $t$ signifies  parametric dependence on time through slow variation of $\mathbf{M}_i(t)$]. The other three contributions---from $\hat{\rho}_\mathrm{geo}(t)$ and $\hat{\rho}_\mathrm{sea}(t)$ governed by the Fermi sea and $\hat{\rho}_\mathrm{surf}(t)$ governed by the Fermi surface electronic  states---contain first nonadiabatic correction~\cite{Thomas2012,Bode2011,Bode2012} proportional to velocity  $\partial \mathbf{M}_i/\partial t$, as well as higher order terms due to $\hat{\rho}_\mathrm{lab}(t)$ being exact. These three contributions define STT out of equilibrium~\cite{Zhang2004,Petrovic2018,Mahfouzi2013}
\begin{equation}\label{eq:stt}
\mathbf{T}_i = J_{sd} \langle \hat{\mathbf{s}}_i \rangle  (t) \times \mathbf{M}_i(t) = \mathbf{T}_i^\mathrm{geo} + \mathbf{T}_i^\mathrm{sea} + \mathbf{T}_i^\mathrm{surf}.
\end{equation}
Each term $\mathbf{T}_i ^\mathrm{geo}$, $\mathbf{T}_i^\mathrm{sea}$, $\mathbf{T}_i^\mathrm{surf}$ can be additionally separated into its own DL and FL components [Fig.~\ref{fig:fig1}(a)], as plotted in Figs.~\ref{fig:fig2} and ~\ref{fig:fig3}. Note that $\mathbf{T}_i ^\mathrm{sea}$ is insignificant in both Figs.~\ref{fig:fig2} and ~\ref{fig:fig3}, so we focus on  $\mathbf{T}_i^\mathrm{geo}$ and  $\mathbf{T}_i^\mathrm{surf}$. 

To gain transparent physical interpretation of $\mathbf{T}_i^\mathrm{geo}$ and  $\mathbf{T}_i^\mathrm{surf}$, we first consider the simplest case~\cite{Bode2012,Stahl2017}---a single $\mathbf{M}_1(t)$ in setup of Fig.~\ref{fig:fig1}(a). The STT contributions as a function of the coupling  $\gamma_c$ to the leads (i.e., reservoirs) are shown in Fig.~\ref{fig:fig2}. We use two different values for $J_{sd}$, where large ratio of $J_{sd}=20$ eV and $\hbar\omega=0.001$ eV is perfect adiabatic limit~\cite{Niu1998,Niu1999,Stahl2017}. Nevertheless, even in this limit and for $\gamma_c \rightarrow 0$  we find  $\mathbf{T}_1^\mathrm{geo} \neq 0$ in Fig.~\ref{fig:fig2}(b) as the only nonzero and {\em purely} FL torque. This is also found in closed system of Ref.~\cite{Stahl2017} where  $\mathbf{T}_1^\mathrm{geo}$ was expressed in terms of the spin Berry curvature. As the quantum system becomes open for $\gamma_c>0$, $\mathbf{T}_1^\mathrm{geo}$ is slightly reduced while  $\mathbf{T}_1^\mathrm{surf}$ emerges with small FL [Fig.~\ref{fig:fig2}(b)] and large DL [Fig.~\ref{fig:fig2}(d)] components. The DL torque $\mathbf{T}_1^\mathrm{surf,DL}$ points toward the $z$-axis and, therefore, enhances the Gilbert damping. In the wide-band approximation~\cite{Bruch2016}, the  self-energy \mbox{$\hat{\Sigma}(E) = -i\Gamma\hat{I}_2$} is energy-independent for $E$ within the bandwidth of the lead, which allows us to obtain analytical expression (at zero temperature)
\begin{equation}\label{eq:geostt}
\mathbf{T}^\mathrm{geo}_1(t) = \frac{\hbar\omega}{2\pi}\bigg[ \pi - 2\tan^{-1}\bigg( \frac{\Gamma}{J_{sd}}\bigg)\bigg]\sin\theta~ \bold{e}_\phi(t).
\end{equation}
Here $\bold{e}_\phi(t) = -\sin\omega t~\bold{e}_x + \cos\omega t~ \bold{e}_y$. Thus, in perfect adiabatic limit, $J_{sd}/\hbar\omega \rightarrow \infty$, or in closed system, $\Gamma \rightarrow 0$, $\mathbf{T}_1^\mathrm{geo}$ is independent of microscopic parameters as expected from its geometric nature~\cite{Wen1988}. The always present $\mathbf{T}_i^\mathrm{geo} \neq 0$ means that electron spin is {\em never} along `adiabatic direction'  $\langle \hat{\mathbf{s}}_i \rangle^\mathrm{ad}_t$.

%
%

Switching to DW [Fig.~\ref{fig:fig1}(c)] embedded into a closed quantum system ($\gamma_c=0)$ shows in Fig.~\ref{fig:fig3}(a)--(d) that  only $\mathbf{T}_i^\mathrm{geo} \neq 0$, which also acquires DL component locally with damping or antidamping action depending on the position of LMM. Upon opening the quantum system ($\gamma_c=\gamma$), Fig.~\ref{fig:fig3}(e)--(h) shows emergence of additional $\mathbf{T}^\mathrm{surf}_i \neq 0$ which, however, becomes negligible [Fig.~\ref{fig:fig3}(f),(h)] in the perfectly adiabatic limit $J_{sd}/\hbar\omega \gg 1$.  At first sight, $\mathbf{T}_i^\mathrm{geo,DL} \neq 0$
violates Berry and Robbins original analysis~\cite{Berry1993} according to which an isolated quantum system, with discrete energy spectrum, cannot exert friction onto the classical system. This apparent contradiction is resolved in Fig.~\ref{fig:fig4}(a) where we show that total $\sum_i \mathbf{T}_i^\mathrm{geo,DL} \equiv 0$ is always zero. Conversely, Fig.~\ref{fig:fig4}(a) confirms that total $\left(\sum \mathbf{T}^\mathrm{surf,DL}_i\right)_z \equiv I_\mathrm{L}^{S_z} + I_\mathrm{R}^{S_z}$ is identical to net spin current pumped into the leads via which  the conduction electrons carry away excess angular momentum of precessing LMMs~\cite{Tserkovnyak2005}. Such identity underlies physical picture where spin current generated by time-dependent magnetization becomes DL torque~\cite{Tserkovnyak2005,Zhang2009}. Note that pumped spin current $I_{i \rightarrow j}^{S_z}$  due to $\hat{\rho}_\mathrm{geo}$ or $\hat{\rho}_\mathrm{sea}$ in Fig.~\ref{fig:fig4}(c) can be nonzero locally, but they sum to zero. The nonuniform pumped spin current due to spatially and time varying magnetization has prompted proposals~\cite{Zhang2009,Kim2012} to amend the LLG equation by adding the corresponding DL torque $\mathbf{M} \times \mathcal{D} \cdot \partial \mathbf{M}/\partial t$ with $3 \times 3$ damping tensor $\mathcal{D}$ whose spatial dependence is given by the so-called spin-motive force (SMF) formula. However, SMF correction was estimated to be small~\cite{Kim2012} in the absence of spin-orbit coupling in the band structure. We confirm its smallness in Fig.~\ref{fig:fig4}(a),(b) for our DW case, but this actually reveals that SMF formula produces incorrectly an order of magnitude smaller torque than obtained from our exact $\hat{\rho}_\mathrm{surf}(t)$. Due to possibly complex~\cite{Bajpai2019a} time and spatial dependence of  $\mathbf{T}_i ^\mathrm{surf}$ and $\mathbf{T}_i^\mathrm{geo}$, the accurate path to incorporate them is offered by self-consistent coupling of electronic DM and LLG calculations, as proposed in Refs.~\cite{Petrovic2018,Bostrom2019,Sayad2015} and in full analogy to how electronic friction is included in nonadiabatic MD~\cite{Lu2019,Thomas2012,Bode2011,Lu2012,Dou2020,Dou2018,Dou2017,Hopjan2018,Chen2019}.

\begin{acknowledgments}
This research was supported in part by the U.S. National Science Foundation (NSF) under Grant No. CHE 1566074. 
\end{acknowledgments}


\end{document}